# Modelling correlated marker effects in genome-wide prediction via Gaussian concentration graph models

Carlos Alberto Martínez[1*], Kshitij Khare[2†], Syed Rahman[2‡], Mauricio A. Elzo[1§]

[1]Department of Animal Sciences
[2]Department of Statistics
University of Florida, Gainesville, FL, USA

## ABSTRACT

In genome-wide prediction, independence of marker allele substitution effects is typically assumed; however, since early stages in the evolution of this technology it has been known that nature points to correlated effects. In statistics, graphical models have been identified as a useful and powerful tool for covariance estimation in high dimensional problems and it is an area that has recently experienced a great expansion. In particular, Gaussian concentration graph models (GCGM) have been widely studied. These are models in which the distribution of a set of random variables, the marker effects in this case, is assumed to be Markov with respect to an undirected graph $G$. In this paper, Bayesian (Bayes G and Bayes G-D) and frequentist (GML-BLUP) methods adapting the theory of GCGM to genome-wide prediction were developed. Different approaches to define the graph $G$ based on domain-specific knowledge were proposed, and two propositions and a corollary establishing conditions to find decomposable graphs were proven. These methods were implemented in small simulated and real datasets. In our simulations, scenarios where correlations among allelic substitution effects were expected to arise due to various causes were considered, and graphs were defined on the basis of physical marker positions. Results showed improvements in correlation between phenotypes and predicted breeding values and accuracies of predicted breeding values when accounting for partially correlated allele substitution effects. Extensions to the multiallelic loci case were described and some possible refinements incorporating more flexible priors in the Bayesian setting were discussed. Our models are promising because they allow incorporation of biological information in the prediction process, and because they are more flexible and general than other models accounting for correlated marker effects that have been proposed previously.

**Key words** Correlated allele substitution effects, genome-enabled prediction, graphical models, sparse covariance estimation.

## 1. INTRODUCTION

A feature shared by different statistical models used in genome-wide prediction (Meuwissen et al., 2001) like the family of hierarchical Bayesian regression models known as the Bayesian alphabet (Gianola et al., 2009; Gianola, 2013) and methods based on mixed model equations known as G-BLUP (VanRaden, 2008) and single step G-BLUP (Aguilar et al., 2011), is that marker allele substitution effects are assumed to be mutually independent. Therefore, a diagonal structure for the covariance matrix of these effects is always assumed. However, it is well known that in many populations, especially livestock and plant populations that have undergone selection, linkage disequilibrium (LD) exists. The presence of LD combined with the interactions among genes and complex interactions between gene products taking place in the metabolism point to correlated marker effects. For example, when several SNP are in LD with the same QTL or group of QTL, their effects could be correlated. Another case where SNP effects are expected to be correlated is when a set of SNP are linked to some QTL whose products interact along metabolic pathways.

---

[*] carlosmn@ufl.edu
[†] kdkhare@ufl.edu
[‡] Shr264@ufl.edu
[§] maelzo@ufl.edu

Thus, there is a need to account for the correlation among marker effects because this is closer to reality and brings advantages like the use of information from correlated marker effects in the prediction of the effect of a given marker, inclusion of biological information in the prediction process and a better knowledge of the covariance structure of marker effects for a particular trait or set of traits. So far, there have been few studies attempting to model a non-diagonal covariance matrix of marker effects in genome-wide prediction and none of them has used an approach based on graphical models. Gianola et al. (2003) proposed Bayesian and frequentist approaches to account for correlated SNP allele substitution effects whereas Yang and Templeman (2012) developed a hierarchical Bayesian model imposing an autoregressive structure on the covariance matrix of these effects.

In statistics, estimation of large covariance matrices when the number of variables is larger than the sample size ("big $p$ small $n$") is a contemporary problem of current interest (Khare and Rajaratnam, 2011; Oh et al., 2016). A growing research area to deal with this complex problem is the use of graphical models to find regularized estimators of covariance or precision matrices using both frequentist and Bayesian methods (Carvalho et al., 2007; Letac and Massan, 2007; Rajaratnam et al., 2008). However, the idea of imposing zeros in the precision matrix is not new; it was proposed by Dempster (1972). Many of these methods share the underlying property of yielding shrinkage estimators (Rajaratnam et al., 2008). Graphical models induce zeros in the covariance or precision matrix, thereby reducing the total number of parameters to be estimated. Models inducing sparsity in the covariance matrix are called covariance graph models, while those inducing sparsity in the precision matrix are called concentration graph models. When the joint distribution of the underlying variables (marker effects in this case) is assumed to be multivariate Gaussian, these zeros translate to appropriate marginal independence assumptions (Gaussian covariance graph models) or conditional independence assumptions (Gaussian concentration graph models).

The patterns of zeros in the covariance or precision matrix can be encoded in terms of an undirected graph $G$, hence the term "graphical models". (Dempster, 1972; Dawid and Lauritzen, 1993; Silva and Ghahramani, 2006; Khare and Rajaratnam, 2011; Ben-David et al., 2015). The nodes of this undirected graph $G$ represent the underlying random variables. For Gaussian concentration graph models (GCGM), variables not sharing an edge in $G$ are conditionally independent given all other variables (Dawid and Lauritzen, 1993; Letac and Massam, 2007). Equivalently, some authors refer to this property as the undirected Markov property with respect to $G$ (Ben-David et al., 2015). Covariance estimation using graph models is an appealing and very promising topic in statistical genomics because it allows accounting for correlation of marker effects in the prediction of genotypic values and phenotypes.

To our knowledge, we are the first to adapt GCGM to account for correlated SNP allele substitution effects in genome-wide prediction. A challenge encountered in this process is the following. The theory of GCGM is developed to estimate the precision matrix of an observable $p$-dimensional random vector using a sample of size $n$. However, estimation of dispersion parameters in genome-wide prediction involve the problem of estimating residual variance(s) and the more challenging problem of estimating the covariance or precision matrix of an unobservable random vector containing SNP effects using a single $n$-dimensional vector of phenotypes as well as genotypes. Thus, the objectives of this study were to introduce the theory of GCGM in genome-wide prediction by developing methods to adapt it to perform sparse estimation of the precision matrix of allele substitution effects, and to evaluate its impact on the accuracy of genome-wide prediction.

## 2. METHODS

This section is organized as follows. Because the theory of GCGM is not widely known by quantitative geneticists, it is briefly presented in subsection 2.1 along with the challenges encountered when adapting it to genome-wide prediction. Subsection 2.2 presents the Bayesian approach to solve the problem (Bayes G and Bayes G-D), whereas an EM algorithm is developed in subsection 2.3 to provide a frequentist solution (GML-BLUP). In subsection 2.4, several approaches to build the graph $G$, some of them leading to decomposable graphs, are presented. Finally, in subsection 2.5, simulated and real datasets used to implement our methods are described.

**2.1 Gaussian Concentration Graph Models**

In this section, frequentist and Bayesian approaches for Gaussian concentration graph models are presented. The off-diagonal entries of the inverse covariance matrix correspond to the conditional covariance between pairs of variables given all other variables; therefore, under a GCGM, assumptions about the structure of this matrix are made and, as explained in the introduction, this structure is represented using a graph $G$ (Dawid and Lauritzen, 1993; Letac and Massam, 2007).

When the pattern of zeros is unknown, i.e., the graph is unknown; the interest is to find the null entries and to estimate the non-null entries of the inverse covariance matrix. This problem is known as model selection (Bickel and Levina 2008; Rajaratnam et al., 2008; Khare et al., 2015). Here, the case of a known graph $G$ is considered. The scenario of known $G$ encompasses situations where either the pattern of zeros of the precision matrix is actually known or domain-specific knowledge permits the definition of $G$. Before discussing concentration graph models, the reader not familiar with basic concepts in graph theory is referred to Appendix A.

*2.1.1 The estimation problem*

The statistical problem is the following. Suppose that $Y_1, Y_2, \ldots, Y_n$ is a set of vectors in $\mathbb{R}^p$ identically and independently distributed $MVN(0, \Sigma)$. The inverse of the covariance matrix $\Sigma$ is usually denoted by $\Omega$. In a GCGM, the target is to estimate $\Omega$ instead of $\Sigma$. Let $G = (V, E)$ be a graph with vertices set $V$ and edges set $E$, notice that $|V| = p$. As mentioned before, the graph $G$ defines the null entries in $\Omega$ and hence it defines the sparsity pattern. The vertices of $G$ represent the set of random variables we are dealing with. Formally, the parameter space is the following cone (Dawid and Lauritzen, 1993; Letac and Massam, 2007):
$\Omega \in \mathbb{P}_G = \{A : A \in \mathbb{P}^+ \text{ and } A_{ij} = 0 \text{ whenever } (i,j) \notin E\}$, where $\mathbb{P}^+$ is the space of positive definite matrices.

*Maximum likelihood estimation*

The negative log-likelihood function has the form:
$$l(\Omega) = c + \frac{n}{2}tr(\Omega S) - \frac{n}{2}\log|\Omega|, \Omega \in \mathbb{P}_G,$$
where $c$ involves all terms not depending on $\Omega$, and $S$ is the sample covariance matrix defined as:
$$S = \frac{1}{n}\sum_{i=1}^{n}(Y_i - \bar{Y})(Y_i - \bar{Y})', \bar{Y} = \frac{1}{n}\sum_{i=1}^{n}Y_i.$$

The negative log-likelihood can be slightly modified to obtain the following objective function whose minimization is equivalent to the minimization of $l(\Omega)$:
$$l^*(\Omega) = tr(\Omega S) - \log|\Omega|, \Omega \in \mathbb{P}_G \tag{2.1}$$

In general, there is no closed form solution. An iterative proportional fitting algorithm developed by Speed and Kiiveri (1986) allows minimizing $l^*(\Omega)$. It is based on a partition of the covariance matrix according to the maximal cliques of $G$.

The properties of the graph have mathematical and statistical consequences on the estimation problem (Letac and Massan, 2007; Khare and Rajaratnam, 2011, Khare and Rajaratnam, 2012). Therefore, it is important to assess certain properties of $G$ because their impact on the estimation problem can be advantageous. In this case, it turns out that when $n$ is greater than the size of the largest clique in $G$, $l(\Omega)$ is strictly convex and consequently it has a unique global minimum. In addition, for decomposable graphs, the MLE has a closed form. Now, some results needed to derive this closed form and are presented; these will also be used in the Bayesian estimation of $\Omega$.

If $\Omega$ is a positive definite matrix, there exists a pair $(L, D)$ such that: $\Omega = LDL'$, $L$ is lower triangular with diagonals equal to one, $D$ is positive diagonal (i.e., $D_{ii} > 0 \ \forall \ i = 1,2,\ldots,p$), and the pair $(L, D)$ is unique. This decomposition is known as modified Cholesky decomposition. Decomposable graphs satisfy the following property (Paulsen et al., 1989). The graph $G = (V, E)$ is decomposable, if and only if, for every $\Omega \in \mathbb{P}_G$ with modified Cholesky decomposition $\Omega = LDL'$ it follows that $L \in \mathcal{L}_G = \{L: L \text{ is lower triangular}, L_{ii} = 1 \ \forall \ 1 \leq i \leq p \text{ and } L_{ij} = 0, i > j, (i,j) \notin E\}$. The implication of this result is that the Cholesky matrix $L$ has (in its lower diagonal elements) the same pattern of zeros as $\Omega$. Now consider the following space: $\mathcal{D} = \{D: D \text{ is diagonal with } D_{ii} > 0 \ \forall \ i = 1,2,\ldots,p\}$. To come up with the closed form for the MLE of $\Omega$ when $G$ is decomposable, the minimization is done in terms of the modified Cholesky decomposition of $\Omega$, that is, we work with the transformation $\Omega \to (L, D)$ that corresponds to a bijection from $\mathbb{P}_G$ to $\mathcal{L}_G \times \mathcal{D}$. Thus, the minimization is done with respect to $(L, D)$. After some algebra, the objective function can be written as:
$$l^*(L, D) = \sum_{i=1}^{p}[D_{ii}(L_{.i}'SL_{.i}) - \log D_{ii}]$$
where $L_{.i}$ is the $i^{th}$ column of $L$. Then, focusing on the $i^{th}$ minimization problem in the above sum and taking into account the structure of $L_{.i}$ we can write:
$$L_{.i}'SL_{.i} = \begin{pmatrix}1 & x_i\end{pmatrix}\begin{pmatrix}S_{ii} & (S_{.i}^{>})' \\ S_{.i}^{>} & S^{>i}\end{pmatrix}\begin{pmatrix}1 \\ x_i\end{pmatrix} \tag{2.2}$$

where $x_i$ is a vector of dimension $v_i$ containing the non-constrained entries of $L_{\cdot i}$ and it is formally defined as $x_i = (L_{ji})_{j>i,(i,j)\in E}$, $S^{>i}$ is a submatrix of the sample covariance matrix built using the vertices greater than $i$ that are connected to vertex $i$ and it is defined as $S^{>i} = (S_{jk})_{j,k>i;(i,j),(i,k)\in E}$, $S_{\cdot i}^{>}$ is a column vector defined as $S_{\cdot i}^{>} = (S_{ji})_{j>i,(i,j)\in E}$. Hence, the function to be minimized for the $i^{th}$ case is:

$$D_{ii}\begin{pmatrix}1 & x_i\end{pmatrix}\begin{pmatrix}S_{ii} & (S_{\cdot i}^{>})' \\ S_{\cdot i}^{>} & S^{>i}\end{pmatrix}\begin{pmatrix}1 \\ x_i\end{pmatrix} - \log D_{ii}, D_{ii} > 0, x_i \in \mathbb{R}^{v_i}.$$

After minimizing with respect to $x_i$, plugging the solution into the objective function and minimizing with respect to $D_{ii}$ we obtain:

$$\hat{x}_i = (S^{>i})^{-1}S_{\cdot i}^{>} \tag{2.3}$$

$$\hat{D}_{ii} = \frac{1}{S_{ii}(S_{\cdot i}^{>})'(S^{>i})^{-1}S_{\cdot i}^{>}} \tag{2.4}$$

By combining $\{\hat{x}_i\}_{i=1}^{p-1}$ and $\{\hat{D}_{ii}\}_{i=1}^{p}$ matrices $\hat{L}$ and $\hat{D}$ can be built and by the invariance property of the MLE (Lehmann and Casella, 1998), the MLE of $\Omega$ can be written as $\hat{\Omega} = \hat{L}\hat{D}\hat{L}'$.

*Bayesian estimation*

In Bayesian estimation of $\Omega$, a family of priors commonly used is known as the Diaconis-Ylvisaker prior (Diaconis and Ylvisaker, 1979). They are appealing because among others, they have the property of yielding linear posterior expectations. The way to construct these priors comes from the following theorem (Diaconis and Ylvisaker, 1979) which is stated here for the sake of completeness.

Consider a natural exponential family

$$f_\theta(x) = \exp(x'\theta - k(\theta))h(x)$$

and a family of priors with a kernel of the form:

$$\tilde{\pi}_{n_0,x_0}(\theta) = \exp(n_0 x_0'\theta - n_0 k(\theta)),$$

now define the collection $\{\tilde{\pi}_{n_0,x_0}\}_{n_0,x_0}$ with $\eta_0, x_0$ such that $\tilde{\pi}_{n_0,x_0}(\theta)$ can be normalized. Under this sort of priors the following properties hold:

1. The posterior has the form $\tilde{\pi}_{n_0+n,\frac{n_0 x_0+n\bar{x}}{n_0+n}}(\theta)$

2. $E\left[\frac{\partial K(\theta)}{\partial \theta}\middle| X_1, X_2, \ldots, X_n\right] = \frac{n_0 x_0+n\bar{x}}{n_0+n}$

3. If $\pi(\theta)$ is any prior such that it is not concentrated at a single point and $E\left[\frac{\partial K(\theta)}{\partial \theta}\middle| X\right] = aX + b$ for some constants $a$ and $b$, then $a \neq 0$ and $\pi(\theta)$ is of the form $ce^{\frac{1}{a}b\theta - \frac{1}{a}(1-a)k(\theta)}$, that is, $\pi(\theta)$ has to be a Diaconis-Ylvisaker prior.

The parameter $x_0$ can be interpreted as the mean of $n_0$ prior observations. Consequently, property 2 indicates that the posterior expectation is a linear combination of the prior and the sample mean. Finally, property 3 states posterior linearity. The general idea when building a Diaconis-Ylvisaker prior is to "imitate the likelihood". Recall that the problem of interest is to estimate the inverse covariance matrix $\Omega$ under multivariate normality, then the likelihood of $\Omega$ is:

$$L(\Omega) = c \exp\left(-\frac{(tr(\Omega nS) - n\log|\Omega|)}{2}\right), \Omega \in \mathbb{P}_G,$$

where $c$ represents terms not depending on $\Omega$. Even though $\Omega \in \mathbb{P}_G$, it can be proven that this is an appropriate density function of the unconstrained (non-zero) elements of $\Omega$. When applying the "imitate the likelihood" principle to this problem, it can be shown that the Diaconis-Ylvisaker prior is:

$$\pi_{(U,\delta)}(\Omega) \propto |\Omega|^{\delta/2} \exp(-tr(\Omega U)/2), U \in \mathbb{P}^+, \delta > 0, \Omega \in \mathbb{P}_G.$$

The shape parameter $\delta$ can be also interpreted as a shrinkage parameter (Rajaratnam et al., 2008). Notice that, if $\Omega \in \mathbb{P}^+$ then we have an inverse Wishart density. However, the matrix $\Omega$ is restricted to $\mathbb{P}_G$ and because of that, this family of priors is known as the G-Wishart (GW) (Roverato, 2000). Under this prior and the likelihood $L(\Omega)$, by the Diaconis-Ylvisaker theorem it follows that the posterior density of $\Omega$ is a GW with parameters $nS + U$ and $n + \delta$, where $S$ is the sample covariance matrix. As in the case of the MLE, it turns out that decomposable graphs have an appealing property. For decomposable graphs, posterior expectations can be obtained in closed form (Letac and Massam, 2007, Rajaratnam et al., 2008) and direct samples from the posterior distribution can be obtained, while for non-decomposable graphs MCMC methods can be used.

In the case of decomposable graphs, in a similar approach to the one used to derive the MLE, instead of working with $\Omega$, its modified Cholesky decomposition $\Omega = LDL'$ is used. As will be shown, this allows direct sampling from the GW distribution. Recall that if the graph $G$ is decomposable then $L \in \mathcal{L}_G$ and $D \in \mathcal{D}$. We already defined the density $\pi_{(U,\delta)}(\Omega)$, but now we are considering a bijection from $\mathbb{P}_G$ to $\mathcal{L}_G \times \mathcal{D}$; therefore, applying standard theory we can compute the density of $L$ and $D$. To this end, the Jacobian of the transformation has to be found. To ease this computation, an ordering strategy of the entries of $\Omega$ is used. The details are omitted, but a brief description is presented. The elements of $L$ and $D$ are ordered as follows:

$$D_1, L_{21}, D_2, L_{31}, L_{32}, \ldots, D_j, L_{(j+1)1}, L_{(j+1)2}, \ldots, L_{(j+1)j}, D_{j+1}, \ldots, D_p,$$

then, the zero entries are ignored. On the other hand, the elements of $\Omega$ are ordered row-wise removing redundancies, that is, $\Omega_{11}, \Omega_{21}, \Omega_{22}, \ldots, \Omega_{p(p-1)}, \Omega_{pp}$ and the zero entries are ignored as well. Recall that the null entries in the inferior off diagonal part of $L$ are the same of $\Omega$. With this ordering strategy, it turns out that the Jacobian is an upper triangular matrix whose diagonal entries are either 1's or the $D_{ii}$ such that $i \in \{i: i > j, (i,j) \in E\} \; \forall \; j = 1,2,\ldots,p-1$. Thus, the determinant of the Jacobian is:

$$|J| = \prod_{j=1}^{p-1} D_j^{n_j},$$

where $n_j = |\{i: i > j, (i,j) \in E\}|$. The operator $|\cdot|$ represents determinant when the argument is a matrix and cardinality when the argument is a set. Therefore, the density on $\mathcal{L}_G \times \mathcal{D}$ induced by $\pi_{(U,\delta)}(\Omega)$ is:

$$\pi_{(U,\delta)}(L,D) \propto \exp\left(\frac{-tr(\Omega U)}{2}\right) \left(\prod_{j=1}^{p-1} D_j^{\frac{\delta+2n_j}{2}}\right) D_p^{\frac{\delta}{2}}.$$

After some algebraic manipulations, this density can be rewritten as:

$$\pi_{(U,\delta)}(L,D) \propto \left(\prod_{j=1}^{p-1} \exp\left(-\frac{c_j D_j}{2}\right) D_j^{(\delta+2n_j)/2} \exp\left(-\frac{D_j}{2}(x_j - a_j)'(U^{>j})(x_j - a_j)\right)\right)$$

$$\times \exp\left(\frac{-U_{pp}D_p}{2}\right) D_p^{\delta/2}$$

where $x_j$ is defined as before, $a_j = -(U^{>j})^{-1}U^{>}_{\cdot j}$, $c_j = U_{jj} - (U^{>}_{\cdot j})'(U^{>j})^{-1}U^{>}_{\cdot j}$, $U_{jj}$ and $U^{>j}$ are the elements resulting from partitioning $U$ in the same way that $S$ was partitioned in (2.2). From $\pi_{(U,\delta)}(L,D)$ it follows that $\{(x_j, D_j)\}_{j=1}^{p-1}$ and $D_p$ are mutually independent random variables, a fact that is advantageous when sampling from this joint distribution. Notice that:

$$\pi_{(U,\delta)}(x_j, D_j) \propto \exp\left(\frac{-c_j D_j}{2}\right) D_j^{\frac{\delta+2n_j}{2}} \exp\left(-\frac{D_j}{2}(x_j - a_j)'(U^{>j})(x_j - a_j)\right), j = 1,2,\ldots,p-1$$

conditioning on $D_j$ yields: $\pi(x_j|D_j) \propto \exp\left(-\frac{D_j}{2}(x_j - a_j)'(U^{>j})(x_j - a_j)\right)$ hence, $x_j|D_j \sim MVN\left(a_j, \frac{1}{D_j}(U^{>j})^{-1}\right)$. Also:

$$\pi(D_j) \propto \int_{\mathbb{R}^{\nu_j}} \exp(-c_j D_j/2) D_j^{(\delta+2n_j)/2} \exp\left(-\frac{D_j}{2}(x_j - a_j)'(U^{>j})(x_j - a_j)\right) dx_j$$

$$\propto \exp(-c_j D_j/2) D_j^{(\delta+|n_j|)/2},$$

consequently, $D_j \sim Gamma\left(\frac{\delta+n_j}{2} + 1, \frac{c_j}{2}\right)$. Similarly,

$$\pi_{(U,\delta)}(D_p) \propto \exp\left(-\frac{U_{pp}D_p}{2}\right) D_p^{\frac{\delta}{2}} \therefore D_p \sim Gamma\left(\frac{\delta}{2} + 1, \frac{U_{pp}}{2}\right).$$

Therefore, $\pi(x_j|D_j)$ and $\pi(D_j)$ are standard distributions and direct sampling can be implemented.

When the graph is non-decomposable, indirect approaches as MCMC methods can be used. One approach is the block-Gibbs sampler proposed by Piccioni (2000) which involves a partition of $\Omega$ according to the set of maximal cliques of the graph $G$. Although this partition is not proper in the sense that it is not disjoint, Piccioni (2000) showed that despite this fact, convergence of the algorithm is attained. Other methods have also being proposed to sample from a GW distribution. Mitsakakis et al. (2011) derived an independent Metropolis algorithm and Lenkoski (2013) came up with a method to obtain direct samples.

### *2.1.2 Challenges encountered when adapting GCGM to genome-wide prediction*

Genome-wide prediction is basically a regression problem. The model considered here is as follows:

$$y = Wg + e \qquad (2.5)$$

where $y$ is an observable $n$-dimensional random vector containing response variables (e.g., corrected phenotypes), $g$ is an unknown $m$-dimensional random vector of marker allele substitution effects, $e$ is a vector of residuals, $W$ is an observable random matrix containing one to one mappings from the set of possible genotypes for every individual at every locus to a subset of the integers. In this study, these mappings are defined as follows:

$$W = \{w_{ij}\} = \begin{cases} 1, & \text{if genotype} = BB \\ 0, & \text{if genotype} = BA \\ -1, & \text{if genotype} = AA \end{cases}$$

where $w_{ij}$ is the mapping of the genotype of the $i^{th}$ individual for the $j^{th}$ marker. The distributional assumptions are the following:

$$\boldsymbol{g}|\Omega \sim MVN(0, \Omega^{-1})$$
$$\boldsymbol{e}|\sigma^2 \sim MVN(0, \sigma^2 I).$$

Under GCGM, the current methods require an independent and identically distributed (IID) sample which is assumed to be drawn from a multivariate normal distribution. The objective is not to estimate the covariance matrix of $\boldsymbol{y}$ (denoted as $V$) but $\Omega$. On the other hand, the residuals are commonly assumed to be homoscedastic and therefore $R$ is a sparse matrix. Sometimes this matrix is assumed to have another structure, but it is generally sparse. Statistical methods for covariance estimation using graphical models have been proposed for the situation in which $V = L + S$ where $L$ is a low rank symmetric matrix and $S$ is a sparse matrix (Zhang et al., 2013). However, these methods require a sample $\boldsymbol{y_1}, \boldsymbol{y_2}, \dots, \boldsymbol{y_k}$ of $n$-dimensional vectors that in this particular case is not available because individuals are typically measured only once for a given trait. For example, the total milk yield of a cow in a given lactation and the weaning weight of a calf are measured only once. In addition, this methodology uses graphical models only to estimate the residual covariance or inverse covariance matrix, not the covariance matrix of regression coefficients (Zhang et al., 2013). Consequently, the theory of GCGM cannot be directly applied to genome-wide prediction and statistical methods have to be developed to adapt it to estimate $\Omega$ and predict marker effects under model 2.5. In the following, Bayesian and frequentist solutions to this problem are presented.

## 2.2 Bayesian approach

Hierarchical Bayesian models suit this problem very well because they provide a principled, natural, and simple way to estimate the inverse covariance matrix of the unobservable vector containing marker allele substitution effects. Given a graph $G$, the following layers are added to complete the hierarchical representation of model 2.5 which is named Bayes G:

$$\sigma^2 \sim Inverse\ Gamma\left(\frac{\tau^2}{2}, \frac{v}{2}\right) \coloneqq IG\left(\frac{\tau^2}{2}, \frac{v}{2}\right)$$
$$\Omega|G \sim GW(\delta, U).$$

Under this prior, the joint posterior is:

$$\pi(\boldsymbol{g}, \sigma^2, \Omega|\boldsymbol{y}, G) \propto f(\boldsymbol{y}|\boldsymbol{g}, \sigma^2, W)\pi(\boldsymbol{g}|\Omega)\pi(\Omega|G)\pi(\sigma^2)$$
$$\propto (\sigma^2)^{-\frac{n}{2}} \exp\left(\frac{-1}{2\sigma^2}(\boldsymbol{y} - W\boldsymbol{g})'(\boldsymbol{y} - W\boldsymbol{g})\right) |\Omega|^{1/2} \exp\left(\frac{-1}{2}\boldsymbol{g}'\Omega\boldsymbol{g}\right)$$
$$\times |\Omega|^{\delta/2} \exp(-tr(\Omega U)/2)\, (\sigma^2)^{-\left(\frac{v}{2}+1\right)} \exp\left(\frac{-\tau^2}{2\sigma^2}\right)$$

As mentioned before, for decomposable graphs, there exists a closed form for the posterior mean under the conventional GCGM. This is possible because the posterior distribution of $\Omega$ is known to be a $GW(\delta + n, U + nS)$, but under the model considered here, this posterior distribution cannot be found in closed form. Therefore, even for decomposable graphs, it is necessary to use an MCMC algorithm to sample from the posterior distribution in order to obtain point estimates of the precision matrix, the residual variance, and the marker effects. To implement a Gibbs sampler, all the full conditionals have to be known so that we can easily sample from them. All the distributions considered here are routinely used in regression models except for the prior for $\Omega$; fortunately, due to the fact that the full conditional density of $\Omega$ under model 2.5 is $\pi(\Omega|Else) = \pi(\Omega|\boldsymbol{g}, G)$ and the conjugacy property of the Diaconis-Ylvisaker prior, it follows that $\Omega|Else \sim GW(\delta + 1, U + \boldsymbol{gg}'), \Omega \in \mathbb{P}_G$, and consequently samples from this full conditional can be obtained using some of the methods mentioned before. This simple but key property allows a straightforward implementation of a Gibbs sampler in Bayes $G$. The remaining full conditionals are standard, but are shown for completeness.

$$\boldsymbol{g}|Else \sim MVN\left(\left(\Omega + \frac{W'W}{\sigma^2}\right)^{-1} \frac{1}{\sigma^2} W'\boldsymbol{y}, \left(\Omega + \frac{W'W}{\sigma^2}\right)^{-1}\right)$$

$$\sigma^2|Else \sim IG\left(\frac{v + n}{2}, \frac{(\boldsymbol{y} - W\boldsymbol{g})'(\boldsymbol{y} - W\boldsymbol{g}) + \tau^2}{2}\right)$$

For general graphs, some of the methods developed to sample from the $GW$ distribution are computationally demanding, and the time to obtain a single sample of the precision matrix of vectors of moderate size (e.g., 1000 to 2000) can make the implementation of a Gibbs sampler computationally intractable. Therefore, it is useful to take advantage of the properties of decomposable graphs whenever possible. Before focusing on the problem of finding decomposable graphs, which is deferred to subsection 2.4, the full conditional distributions of the Cholesky parameters of $\Omega$ given that $G$ is decomposable are presented. Using the forms of the distributions of $\boldsymbol{x}_j|D_j, j = 1,2,\ldots,m-1$ and $D_j, j = 1,2,\ldots,m$, presented in subsection 2.1.1, it follows that under model 2.5 $\boldsymbol{x}_1, \ldots, \boldsymbol{x}_{m-1}|Else$ distribute independently as:

$$\boldsymbol{x}_j|Else \sim MVN\left((U^{>j} + (\boldsymbol{gg}')^{>j})^{-1}(U_{\cdot j}^{>} + (\boldsymbol{gg}')_{\cdot j}^{>}), (U^{>j} + (\boldsymbol{gg}')^{>j})^{-1}\right)$$

where matrices $U$ and $\boldsymbol{gg}'$ are partitioned as in 2.2. Furthermore $D_1, \ldots, D_m|Else$ distribute independently as:

$$D_j|Else \sim Gamma\left(\frac{\delta + n + n_j}{2} + 1, \frac{(\boldsymbol{gg}')_{jj} + U_{jj} - Q(U, \boldsymbol{g})}{2}\right), j = 1,2,\ldots,m-1$$

$$Q(U, \boldsymbol{g}) = (U_{\cdot j}^{>} + (\boldsymbol{gg}')_{\cdot j}^{>})'(U^{>j} + (\boldsymbol{gg}')^{>j})^{-1}(U_{\cdot j}^{>} + (\boldsymbol{gg}')_{\cdot j}^{>})$$

$$D_m|Else \sim Gamma\left(\frac{\delta + n}{2} + 1, \frac{(\boldsymbol{gg}')_{mm} + U_{mm}}{2}\right).$$

This particular case of our hierarchical Bayesian model arising when $G$ is decomposable is referred to as Bayes G-D. Sampling from these distributions at iteration $t$ permits to build matrices $L^{(t)}$ and $D^{(t)}$ which in turn allow obtaining the precision matrix as $\Omega^{(t)} = L^{(t)}D^{(t)}(L^{(t)})'$.

## 2.3 Frequentist approach

Unlike the Bayesian approach, envisaging a frequentist solution to the problem of adapting GCGM to genome-wide prediction is not straightforward and there is not a direct and principled method to cope with this problem. Therefore, some *ad hoc* assumptions have to be done in order to provide a frequentist formulation. The method proposed here is named Graphical Maximum

Likelihood-BLUP (GML-BLUP) because it involves two steps, combining the EM algorithm (Dempster et al., 1977) with GCGM to estimate covariance components (an algorithm denoted as G-EM) and plugging this estimates into mixed model equations corresponding to model 2.5 to obtain the empirical BLUP of $g$ (Henderson, 1950; 1963). According to the theory of GCGM presented in subsection 2.1.1, the objective function 2.1 is strictly convex if the number of observations is greater than the size of the largest clique of $G$ denoted as $C$. To meet this requirement under the rationale of the EM algorithm, the following approach is used. Suppose that data can be split into $f > C$ groups such that each group has a different vector of marker effects, that is, $y_i = W_i g_i + e_i, \forall\ i = 1,2,\dots,f$. It is also assumed that: $g_1, \dots, g_f$ are iid $MVN(0, \Omega^{-1})$, $e_1, \dots, e_f$ are iid $MVN(0, \sigma^2 I_{n_i})$ and $Cov(g_i, e_{i'}) = 0, \forall\ 1 \leq i, i' \leq f$, where $n_i$ is the number of observations in group $i$; therefore, $\sum_{i=1}^{f} n_i = n$. A natural criterion to split data is to cluster individuals in half-sib or full-sib families as is commonly done in quantitative genetics (Falconer and Mackay, 1996). The assumption of heterogeneity of marker effects across families was also considered by Gianola et al. (2003). Under these assumptions, the complete log-likelihood can be written as:

$$l(\sigma^2, \Omega) = constants - \frac{n}{2}\log\sigma^2 + \frac{f}{2}\left(\log|\Omega| - tr(\Omega S_g)\right) - \frac{\|y - W^* g^*\|_2^2}{2\sigma^2} \quad (2.7)$$

$$S_g = \frac{1}{f}\sum_{i=1}^{f} g_i g_i',\ g^* \coloneqq (g_1' \cdots g_f')',\ W^* = Block\ Diag.\{W_i\}_{i=1}^{n}.$$

Following Searle et al. (1992), the expected values of sufficient statistics for the covariance parameters taken with respect to the conditional distribution of the missing data given the observed data have to be found. From 2.6 it follows that the sufficient statistic for $(\Omega, \sigma^2) \coloneqq \theta$ is $(S_g, e^{*\prime} e^*), e^* = y - W^* g^*$. It is easily shown that given $y$, $g_1, \dots, g_f$ are independent with the following distributions: $g_i | y_i \sim MVN\left(K_i^{-1} \frac{W_i' y_i}{\sigma^2}, K_i^{-1}\right)$, where $K_i \coloneqq \frac{W_i' W_i}{\sigma^2} + \Omega$. Similarly, it follows that $e^* | y \sim MVN(\sigma^2 V^{-1} y, \sigma^2 (I - \sigma^2 V^{-1}))$, where $V = W^{*\prime} I_f \otimes \Omega^{-1} W^* + R$. Hence,

$$E[S_g | y] = \frac{1}{f}\sum_{i=1}^{f} K_i^{-1}\left[I_m + \frac{1}{(\sigma^2)^2} W_i' y_i y_i' W_i K_i^{-1}\right] \quad (2.7)$$

$$E[e^{*\prime} e^* | y] = \sigma^2 (n - \sigma^2 tr(V^{-1}) + \sigma^2 y' V^{-1} V^{-1} y) \quad (2.8)$$

Applying the Woodbury's identity, $E[S_g | y]$ can be alternatively expressed as:

$$E[S_g | y] = \frac{1}{f}\Omega^{-1}\left\{f I_m - \left[\sum_{i=1}^{f} W_i' V_i^{-1}(I_{n_i} - y_i y_i' V_i^{-1}) W_i\right]\Omega^{-1}\right\} \quad (2.9)$$

where $V_i \coloneqq W_i \Omega^{-1} W_i' + \sigma^2 I_{n_i}$.

In 2.7, $f\ m \times m$ matrices have to be inverted and these matrices are dense because $W_i' W_i$ is dense. On the other hand, in 2.9, $\Omega$, which is sparse in GCGM, has to be inverted along with $f\ n_i \times n_i$ matrices. The expectation step of the G-EM algorithm consists of using either 2.7 or 2.9 to compute $E[S_g | y]$ and 2.8 to compute $E[e^{*\prime} e^* | y]$, the maximization step is the one involving GCGM. At iteration $t$, the maximization step involves the following computations:

$$(\hat{\sigma}^2)^{(t+1)} = \frac{\hat{q}^{(t)}}{n},\ \hat{q}^{(t)} \coloneqq E[e^{*\prime} e^* | y]\bigg|_{\theta = \theta^{(t)}}$$

$$\widehat{\Omega}^{(t+1)} = h\left(\widehat{S}_g^{(t)}\right), \widehat{S}_g^{(t)} := E[S_g|\mathbf{y}]\big|_{\boldsymbol{\theta}=\boldsymbol{\theta}^{(t)}}$$

where function $h(\cdot)$ is as defined in subsection 2.1.1, it involves evaluating expressions 2.3 and 2.4 taking $\widehat{S}_g^{(t)}$ as argument. Subsequently, matrices $\widehat{L}^{(t+1)}$ and $\widehat{D}^{(t+1)}$ are built and $\widehat{\Omega}^{(t+1)}$ is computed as $\widehat{L}^{(t+1)}\widehat{D}^{(t+1)}\widehat{L}'^{(t+1)}$. Once the algorithm converges and the maximum likelihood estimates of $\Omega$ and $\sigma^2$ are obtained, these are plugged in the mixed model equations corresponding to model 2.5 to obtain the empirical BLUP of $\boldsymbol{g}$ (Henderson, 1950; 1963):

$$\widehat{\boldsymbol{g}} = \left(W'W + \widehat{\sigma}^2\widehat{\Omega}\right)^{-1}W'\mathbf{y}.$$

### 2.4 Some criteria to define $G$ and to order markers

A graph defining the partial correlation structure can be specified according to several genetic criteria. These criteria come from the assumptions about how marker allele substitution effects could be partially correlated. In the following, some alternatives to define $G$ based on genetic and biological principles are discussed.

1. The first approach is based on the idea that there exists spatial correlation (Gianola et al. 2003, Yang and Templeman 2012). In this case, the physical/linkage map of the considered loci is used to define $G$. One simple alternative is to define a window on the basis of a given physical/linkage distance or a given number of markers and slide it across each chromosome. The ordering of markers is the natural ordering imposed by the physical/linkage map. If markers are not equally distanced, for each chromosome it induces a differentially banded precision matrix in the first case and a banded matrix in the second case. Then the precision matrix for all marker effects is the direct sum of banded or differentially banded matrices (one per each chromosome).

2. A second approach to define the groups of loci whose effects are correlated is to use existing gene networks built using the same set of loci being considered. This would induce a graph $G$ with a general structure, general in the sense that *a priori* it is not expected to follow a particular pattern. In this case, the ordering can be also dictated by spatial location of the loci.

3. A third approach is bringing biological information using bioinformatics tools to group the loci according to their function and/or the metabolic pathways in which they are known to be involved. This will create "functional groups" or "functional blocks" of loci, and there are two ways to define the graph.

   3.a The allelic substitution effects of loci in the same group are partially correlated with each other. Correlation among effects of loci in different groups is not allowed. If the ordering is such that groups are ordered first and then markers are ordered within group such that markers in groups with smaller index have smaller indexes, it induces a block-diagonal precision matrix.

   3.b As in 3.a, but correlations among loci in different groups are allowed. The ordering is discussed later.

4. Lastly, some of the existing metrics to assess LD between loci can be used to define $G$ by imposing a threshold and those pairs of loci having a metric greater than the threshold will be assumed to have partially correlated effects.

Because of the aforementioned properties, decomposable graphs have been object of research and whenever possible, it is desirable to work with this kind of graphs. Thus, some rules to define $G$ that guarantee decomposability are presented as well. The following proposition establishes which approaches will induce decomposable graphs. Hereinafter, the "functional blocks" mentioned in approach 3 will be referred to as blocks. If a block contains a subset of markers with effects correlated with the effects of a subset of markers in another block, these blocks are said to be linked. Let $B$ be the total number of blocks and $\mathcal{L}$ be the set of pairs of linked blocks. Let $\Psi$ be the set of blocks linked with at least two other blocks, $\forall\, l \in \Psi$ let $\Gamma_l$ be the set of blocks linked to block $l$ and $\forall\, a \in \Gamma_l$, let $C_{l_a}$ be the subset of markers in block $l$ whose effects are correlated with effects of a subset of markers in block $a$, $1 \leq a \leq B, a \neq l$.

*Proposition 1* The graphs induced under approaches 1 and 3.a are decomposable and the graph induced under 3.b is decomposable if there exists an ordering of markers $\sigma'$ that along with the edges set satisfy the following conditions.

**Condition 1.1** For all possible triplets of linked blocks $\{l, l', l''\}$ such that $C_{l_{l'}} \neq C_{l_{l''}}$, $C_{l'_l} \neq C_{l'_{l''}}$, $C_{l''_l} \neq C_{l''_{l'}}$, and the sets $I_l := C_{l_{l'}} \cap C_{l_{l''}}$, $I_{l'} := C_{l'_l} \cap C_{l'_{l''}}$ and $I_{l''} := C_{l''_l} \cap C_{l''_{l'}}$, are all non-empty, the following never happens: $\sigma'(i) > \sigma'(j) > \sigma'(k)$, $i \in C_{l_{l''}} \cap I_l^C, j \in C_{l'_l}$ or $i \in C_{l_{l''}}, j \in C_{l'_l} \cap I_{l'}^C$, and $k \in I_{l''}$; if there are triplets of linked blocks $\{l, l', l''\}$ such that exactly one of the three sets $\{I_l, I_{l'}, I_{l''}\}$, say $I_l$ is empty, then: $\min\{\sigma'(k), \sigma'(i), \sigma'(j)\} = \sigma'(k), \forall\, k \in C_{l_{l'}} \cup C_{l_{l''}}\, \forall\, j \in I_{l'}\, \forall\, i \in I_{l''}$ and if exactly two of these sets, say $\{I_l, I_{l'}\}$ are empty, then for either $l$ or $l'$, say $l$, $\sigma'(k) < \sigma'(i)\, \forall k \in C_{l_{l'}} \cup C_{l_{l''}}, \forall i \in I_{l''}$. Superindex $C$ indicates the complement with respect to the index set of the corresponding block.

**Condition 1.2** For every possible triplet of blocks $\{l, l', l''\}$ the following does not happen: $\sigma'(k) < \sigma'(j) < \sigma'(i), k \in I_l, j \in C_{l'_l}, i \in C_{l''_l}, C_{l'_{l''}} = \emptyset$.

**Condition 1.3** For every duplet of linked blocks $\{l, l'\}$ the following does not hold: $\exists\, i \in l, \{j, k\} \in l'$ such that $\sigma'(i) > \sigma'(j) > \sigma'(k), i \in C_{l_{l'}}, j \in C_{l'_l}^C, k \in C_{l'_l}$.

**Condition 1.4** For each pair of linked blocks $(l, l')$, $C_{l_{l'}} \times C_{l'_l} \in E_\sigma$, that is, the effect of each marker in $C_{l_{l'}}$ is correlated with the effects of all marker in $C_{l'_l}$.

Moreover, conditions 1.1, 1.2 and 1.3 are necessary whereas condition 1.4 is not.

**Proof** See appendix B.

This proposition involves all possible orderings of markers. However, if markers are ordered in such a way that markers in the same block are given consecutive indices, the number of possible orderings is reduced. Thus, in order to provide a simpler way to order markers, the following proposition only requires the existence of an ordering of the blocks and a structure on the edges set satisfying certain conditions that permit to find a perfect elimination ordering of markers.

*Proposition 2* If there exists an ordering $\rho$ of the blocks which coupled with the structure of the edges set satisfy condition 1.4 plus the following conditions:

**Condition 2.1** $C_{l_a} = \cdots = C_{l_m} := C_l\, \forall\, l \in \Psi$

**Condition 2.2** For every possible triplet of blocks $\{l, l', l''\}$ the following does not happen: $(l, l'), (l, l'') \in L, (l', l'') \notin L, \rho(l) < \rho(l') < \rho(l'')$.

Then the following ordering strategy (denoted by $\sigma$) of marker loci is a perfect elimination ordering: once blocks have been ordered according to $\rho$, markers are ordered in such a way that the smaller the index of a block the smaller the indices of the markers pertaining to that block. The ordering inside each block is done as follows: markers in $C_l$ are given the largest indices in block $l$. In addition, under this ordering strategy, condition 2.2 is also necessary for $\sigma$ to be a perfect elimination ordering whereas condition 2.1 is not.

**Proof** See Appendix C.

**Corollary to Proposition 2** Consider the "super graph" formed by regarding the blocks as super nodes and $\mathcal{L}$ as a "super vertices set". Then, under conditions 2.1 and 1.4, if the "super vertices set" admits a perfect elimination ordering, the ordering defined in proposition 2 corresponds to a perfect elimination scheme.

**Proof** See Appendix D.

### 2.5 Simulated and real data analyses

To ensure computational tractability, a small genome comprised of a single chromosome of 1 M length, with 1000 biallelic QTL and 3201 SNP markers was simulated through a forward-in-time approach using the software QMSim (Sargolzaei and Schenkel, 2013). A historical population was simulated by creating 1000 generations of random mating in order to reach mutation-drift equilibrium and to create LD (Sargolzaei and Schenkel, 2013). Each generation of the historical population consisted of 500 males and 500 females. The phenotypes were simulated as the sum of breeding values and an error term. Errors were drawn from independent normal distributions and the heritability of the trait was 0.5. Two strategies to simulate the QTL allelic effects were considered. The first one involved sampling the effects from independent $N(0,1)$ distributions whereas in the second one QTL effects were drawn from a $MVN(0, M)$ distribution, where $M^{-1}$ was a banded diagonal matrix with band size 10. In both cases, QTL effects were scaled to obtain an additive genetic variance of 50 (strategy 1) and 100 (strategy 2). The same QTL and SNP genotypes were used in both strategies. Simulation of errors, and QTL effects for strategy 2 as well as all the statistical analyses were performed using existing R packages and in-house R scripts (R Core Team, 2015). Ten replicates of each simulation were obtained and only decomposable graphs were considered. Markers included in the analyses were those with minor allele frequency greater or equal than 0.01. The graph $G$ was built using approach 1, considering both, windows defined by a fixed number of marker loci (6) or a fixed map distance (0.1875 cM). When windows were based on number of markers models were denoted as Bayes G-D_B and GML-BLUP_B and when windows were defined by a given map distance models were denoted as Bayes G-D_DB and GML-BLUP_B. These window sizes were defined on the basis of the average LD decay, considering the distance at which the average LD was smaller than 0.1. Also, some conventional models used in genome-wide prediction were fitted; Bayes A and Bayesian Ridge Regression (BRR). Three sets of analyses were performed. The first one corresponded to fitting models considering SNP effects to datasets simulated with strategy 1 (scenario 1). In the second one, the same models were fitted to datasets from simulation strategy 2 (scenario 2). Finally, the third set involved fitting our models to datasets from strategy 2, but models considered QTL instead of SNP effects (scenario 3, the ideal scenario).

In each replicate, the training set was defined as generations 0,1 and 2, and generation 3 was the validation set.

Real data analyses were performed for three different traits using a small dataset from the University of Florida's Beef Research Unit (BRU). Briefly, the three phenotypes were daily feed intake (DFI), the ultrasonic measure of percent of intra-muscular fat (UPIMF) and live weight taken the same day than UPIMG (UW). These records came from individuals genotyped with the Illumina GoldenGate Bovine3K BeadChip (Illumina, Inc., 2011) belonging to a multibreed Angus-Brahman herd. For details on the multibreed herd, the genotypes and the traits see Elzo et al. (2012) and Elzo et al. (2013). The heritabilities of these traits estimated in the multibreed herd were 0.31 for DFI (Elzo et al., 2012), 0.54 for UW and 0.53 for UPIMF (Elzo et al., 2013). The same models fitted to simulated data except those with differentially banded structure were fitted to these traits after correcting data for the following fixed effects: contemporary group (year-pen), age of dam, age of individual, sex, Brahman fraction and heterozygosity of the individual. After editing data such that only individuals with complete genotypes were included, the final dataset had 102 individuals and 2407 SNP. The training set included the oldest 72 individuals, and the validation set was composed by the 30 remaining individuals. In this case, the windows used to define $G$ (which determine the bandwidth of the precision matrix) were defined such that they contained 5 adjacent markers or based on observed LD. In the latter case, for each chromosome, the number of SNP defining the bandwidth of the precision matrix was chosen using the average distance for which the LD was larger than 0.1. Genotypes for denser chips are available in this herd, but this small dataset was used to ensure computational tractability. Only within chromosome correlations were allowed.

In all frequentist analyses, individuals were grouped according to half-sib families. Models were compared through the Pearson correlation of phenotypes and predicted breeding values in the validation set (hereinafter predictive ability) for real and simulated data, and the Pearson correlation between true and predicted breeding values (accuracy) in training and validation sets for simulated data. For Bayesian models, a total of 25000 MCMC samples were obtained using the Gibbs sampler described in subsection 2.2 considering the first 10000 samples as burn in. For the GML-BLUP models, convergence was declared when the change in the average absolute difference of the parameters in two consecutive iterations was smaller than $1\times 10^{-4}$.

## 3. RESULTS AND DISCUSSION

In this study, Bayesian and frequentist methods to consider partially correlated marker effects in genome-wide prediction using GCGM were developed. Several approaches, based on genetic principles, to define the graph $G$ were proposed and it was shown that some of these approaches induced a decomposable graph (some of them requiring particular conditions). The performance of the proposed models was addressed in small simulated and real datasets. Two of the conventional hierarchical Bayesian models used in genome-wide prediction (Bayes A and BRR) were also fitted to these data. The simulated datasets were composed by two sets of ten replicates each. The two sets shared the simulated QTL and SNP genotypes. In the first set, QTL effects were sampled from independent normal distributions, while in the second set these were sampled from a multivariate Gaussian distribution with a banded precision matrix. Analyses were performed under three scenarios using these sets of simulated data. In scenario 1, it was expected that correlations arose from physical proximity to causal variants. On the other hand, in scenario 2, the fact that QTL effects were correlated introduced another potential source of correlation among SNP effects. Lastly, in scenario 3, phenotypes were regressed on QTL genotypes instead of SNP genotypes; thus, under this ideal scenario it was expected *a priori* that our models would outperform standard models assuming independence. It is said that this is an ideal scenario because the causal variants are considered in the model instead of proxies like molecular markers.

### 3.1 Predictive abilities and accuracies in simulated and real datasets

Table 1 shows the summary of the performance of the different models considered here in simulated datasets under scenarios 1, 2 and 3.

**Table 1** Average (over 10 replicates) predictive abilities (APA), accuracies in training (AAT) and validation (AAV) sets for simulated datasets. Standard deviations are presented inside the brackets.

| Model | Scenario 1 | | | Scenario 2 | | | Scenario 3 | | |
|---|---|---|---|---|---|---|---|---|---|
| | APA | AAT | AAV | APA | AAT | AAV | APA | AAT | AAV |
| Bayes G-D_B | 0.428 (0.058) | 0.750 (0.032) | 0.588 (0.090) | 0.386 (0.080) | 0.740 (0.025) | 0.556 (0.081) | 0.512 (0.055) | 0.791 (0.025) | 0.710 (0.043) |
| Bayes G-D_DB | 0.424 (0.061) | 0.750 (0.028) | 0.587 (0.086) | 0.380 (0.083) | 0.737 (0.023) | 0.537 (0.081) | -- | -- | -- |
| GML-BLUP_B | 0.540 (0.065) | 0.805 (0.033) | 0.730 (0.068) | 0.536 (0.054) | 0.800 (0.027) | 0.728 (0.037) | 0.577 (0.038) | 0.830 (0.025) | 0.771 (0.028) |
| GML-BLUP_DB | 0.530 (0.062) | 0.804 (0.030) | 0.722 (0.071) | 0.547 (0.050) | 0.810 (0.026) | 0.744 (0.036) | -- | -- | -- |
| Bayes A | 0.341 (0.054) | 0.736 (0.025) | 0.463 (0.081) | 0.304 (0.101) | 0.720 (0.024) | 0.440 (0.092) | 0.478 (0.038) | 0.735 (0.017) | 0.630 (0.071) |
| BRR | 0.453 (0.062) | 0.764 (0.027) | 0.616 (0.096) | 0.439 (0.069) | 0.739 (0.024) | 0.619 (0.071) | 0.504 (0.046) | 0.780 (0.026) | 0.706 (0.044) |

In scenario 1, APA, AAT and AAV were higher for GML-BLUP models, whereas Bayes A had the poorest performance; on the other hand, BRR and Bayes G-D performed similarly. Likewise, in scenario 2, the GML-BLUP models had the best predictive performance and Bayes A had the worst; however, this time the difference between Bayes G-D and BRR was more marked, the later model outperformed Bayesian models accounting for correlated effects. Finally, in scenario 3, the performance of all models was more similar than in the previous cases. Again, GML-BLUP had the highest APA, AAT and AAV and Bayes A had the lowest, but this time Bayes G-D slightly outperformed BRR. In general, Bayes A showed the worst performance and GML-BLUP the best one, and there were tight differences between models considering banded precision matrices and those considering differentially banded precision matrices. Perhaps, the reason for the superiority of the frequentist models is the use of family information when estimating $\Omega$. On the other hand, Bayes A assumes independent marker effects, each one having a different variance, and for these datasets this assumption yielded the worst results. When moving from scenario 1 to scenario 2, there was not an increment in the overall performance of models accounting for partial correlation, in fact, for Bayesian models there was a light drop in APA, AAT and AAV. A potential explanation is the fact that the bandwidths of the precision matrix were defined on the basis of LD between pairs of markers and were not tuned. Tuning was not performed due to limited computational resources. Recall that the concentration matrix of marker effects was unknown, because QTL instead of SNP effects were simulated. Consequently, misspecification of the partial correlation structure may have been another possible cause of this result, i.e., perhaps the partial correlation matrix of SNP effects was not banded. When simulated QTL effects were partially correlated and models were parameterized in terms of these effects (scenario 3), all models accounting for partially correlated effects performed better than Bayes A and BRR; however, in terms of APA, differences between Bayes G-D and BRR were small. As expected, the performance of all models improved markedly when fitting QTL instead of marker effects, the increment being more visible for Bayes G-D. In these simulations, the effect of factors such as number of QTL controlling the trait, number of SNP, and heritability of the trait were not explored. Regarding heritability, a trait with low heritability (0.01) was simulated following strategy 1 described before, then, Bayes G-D_B, GML-BLUP_B, Bayes A and BRR were fitted using mappings of SNP genotypes as explanatory variables. In this case, Bayes G-D_B had the highest APA, AAT and AVV and GML-BLUP_B did not show the superiority exhibited for the high heritability trait. Table 2 summarizes the results of this analysis.

**Table 2** Model performance for a low heritability trait, standard deviations are shown inside the brackets.

| Model | APA | AAT | AAV |
|---|---|---|---|
| Bayes G-D_B | -0.008 (0.097) | 0.208 (0.082) | 0.163 (0.0608) |
| GML-BLUP_B | -0.011 (0.085) | 0.192 (0.069) | 0.132 (0.104) |
| Bayes A | -0.011 (0.092) | 0.187 (0.116) | 0.144 (0.082) |
| BRR | -0.015 (0.089) | 0.187 (0.100) | 0.156 (0.076) |

These results suggest that the relative performance of these set of models could change with the heritability of the trait and therefore, it is an issue that has to be studied in future research. Due to the low heritability of this trait, there is little or null benefit from the use of family information in GML-BLUP and this is what was observed in this analysis. Table 3 contains the predictive abilities for the three traits in the real dataset.

**Table 3** Predictive abilities for real data analyses

| Model[1] | DFI | UPIMF | UW |
|---|---|---|---|
| Bayes G-D_B | 0.2 | -0.17 | 0.03 |
| Bayes G-D_B* | 0.1 | 0.04 | -0.04 |
| GML-BLUP_B | 0.02 | 0.09 | -0.06 |
| GML-BLUP_B* | -0.02 | 0.03 | -0.04 |
| Bayes A | 0.08 | -0.06 | -0.007 |
| BRR | -0.18 | -0.19 | -0.17 |

[1]The asterisk indicates that the bandwidth of the precision matrix was defined on the basis of LD.

In this real dataset, for DFI, Bayes G-D assuming a banded matrix with a bandwidth of 5 yielded the highest predictive ability and BRR the lowest. Bayes G-D assuming a bandwidth defined on the basis of LD had the second highest predictive ability. For UPIMF, GML-BLUP_B had the best performance, followed by Bayes G-D_B* and GML-BLUP_B*, model BRR had the lowest. Finally, for UW, the trait with the lowest predictive abilities, Bayes G-D_B outperformed the other models which had similar predictive abilities, all negative and close to zero. For DFI, predictive abilities from the Bayesian models proposed in this study were higher than those obtained by Elzo et al. (2012) for the same population, but using a larger sample size, a higher number of markers and models that considered: only marker effects, marker and polygenic effects, and only polygenic effects. Albeit the estimated heritabilities of these traits were from moderate to high, predictive abilities were low. This could be the result of the low marker density and the small sample size. Although these results need to be validated in a larger population and using a higher number of markers, they suggest a superiority of models developed here over models assuming independent marker allele substitution effects as well as a marked difference in the performance of models assuming different bandwidths of the precision matrices. Model BRR always yielded negative predictive abilities.

Yang and Templeman (2012) found increments in accuracy of breeding values in the testing population up to 3% in simulated datasets and 3.6% in a real dataset when comparing their first order antedependence models with their independent marker effects counterparts Bayes A and Bayes B. The simulations carried out in Yang and Templeman (2012) are closer to those considered in scenario 1, where it is expected that correlation among SNP arises from physical proximity to the causal variants. The heritability used in their simulations was also 0.5 and they also simulated a genome formed by a single chromosome of 1 M length. In this study, differences in accuracy in the testing population in scenario 1 were larger than those found by Yang and Templeman (2012). The ratio of number of SNP to number of QTL was 3.2, perhaps it was too low for SNP effects to be correlated due to linkage with the same causal variants. In Yang and Templeman (2012) these ratios ranged from 3.2 to 72.73, and they considered only 30 QTL. Yang and Templeman (2012) reported improvements in accuracy in simulated and real datasets when comparing the antedependence models with their counterparts Bayes A and Bayes B; however, their results showed that Bayes B outperformed ante-Bayes A in terms of accuracy for simulated data and predictive ability for the mice data set described in Legarra et al. (2008). The authors attributed the observed results for simulated data to the small number of simulated QTL. The study of Gianola et al. (2003) did not consider data analyses.

## 3.2 Some comments about the models

Our models account for partially correlated SNP effects through imposing the Markov property (Dawid and Lauritzen, 1993; Letac and Massam, 2007) with respect to $G$ over these random variables. The graph is defined based on some biological premises. This specification might be crucial for the predictive performance of these models and it is an aspect that we did not explore in this paper that poses a very interesting topic for further research. Notice that the conventional genome-wide regression models assuming a diagonal covariance matrix, and therefore a diagonal precision matrix as well; are also imposing a conditional covariance structure. This corresponds to the especial case when $E = \emptyset$, that is, the edges set is the empty set. An extension to be considered in future studies is to use the richer family of priors developed by Letac and Massam (2007) which is known as the type II Wishart family. It is more flexible as it permits to include a $k+1$-dimensional shape parameter where $k$ is the number of cliques of $G$. It allows inducing differential shrinkage trough the precision matrix. Recall that the priors used in this study have a single shape or shrinkage parameter $\delta$. The only limitation of the type II Wishart family of priors is that it is restricted to decomposable graphs; nevertheless, recent research by Ben-David et al. (2015) has extended this flexible family of priors for the case of general DAG's and their family is even more flexible because it has as many shape parameters as the DAG has nodes.

As noticed in subsection 2.3, formulating a frequentist solution to the problem considered in this study is not as straightforward as doing it from the Bayesian perspective. Hence, this is another example of a problem that can be easily solved using the Bayesian paradigm, but whose frequentist solution is troublesome. Taking advantage of hierarchical models, which allow modeling complex processes (Lehmann and Casella, 1998), it was possible to provide a principled and simple solution from the Bayesian side. On the other hand, in order to find a frequentist solution, it was necessary to develop an *ad hoc* procedure that involves a partition of data which permits the computation of the MLE of the precision matrix. It is based on the assumption that different families have different sets of marker effects. It is necessary because otherwise function 2.1, which is used in the M step of the G-EM algorithm, is not strictly convex and therefore a global minimum does not exist. Under the assumption that $\boldsymbol{g}$ is a random, the problem is to estimate the realization of this random vector that determined the observed data (Henderson, 1984). If there is a single vector $\boldsymbol{g}$, it is assumed that this realization is the same for the whole population. However, if marker allele substitution effects are random, it could be possible that the realization varies across certain groups or even across individuals. When formulating a mixed effects model to account for correlated marker effects, Gianola et al. (2003) also considered the possibility of having different marker affects across families. In the GML-BLUP approach, this idea is only used to estimate $\Omega$ because later on, this MLE is plugged in the mixed model equations corresponding to model 2.5 which assumes a single $\boldsymbol{g}$. From a pragmatic point of view, the assumptions of heterogeneous marker effects across families can be seen as a trick and thinking of this GML-BLUP procedure as a predictive machine, it is worth to assess its predictive performance, which indeed was satisfactory in the small datasets considered here. Of course, empirical BLUP for family-specific vectors $\boldsymbol{g}_1, \ldots, \boldsymbol{g}_f$ under the assumptions presented in subsection 2.3 can be easily obtained using the appropriate mixed models equations and it could be addressed if the accuracy and predictive ability of this alternative approach are better.

When using mixed model methodologies to compute the BLUP of marker effects, the inverse covariance matrix has to be found in order to build the mixed model equations. In the classical animal model, the so-called Henderson rules (Henderson, 1976) allowed an efficient computation of this matrix. In the case of a model parameterized in terms of markers as in VanRaden (2008), finding the inverse covariance matrix is not an issue because it is diagonal. When allowing partially correlated effects, this matrix is no longer diagonal, and although it is sparse, there are no rules to avoid direct inversion; however, an advantage of the GEM-BLUP method is that the inverse covariance matrix is obtained directly, so there is no need for inversion. This feature can be seen as another advantage of using GCGM in genome-wide prediction.

### 3.3 Building the graph $G$

The criteria to define $G$ presented in subsection 2.4 may not reflect the true underlying conditional covariance structure; notwithstanding, it could capture some partial correlations that might improve the accuracy of genome-wide prediction. Of course, an ideal scenario would be to identify the real underlying conditional covariance structure which would be expected to improve the predictive performance of the models. At this point, it is worth to mention methods that "let the data talk" to define $G$. These methods exist in the theory of GCGM and as mentioned in the introduction, this is a model selection problem where the sparsity pattern of the precision matrix has to be found and the non-zero entries have to be estimated (Bickel and Levina 2008; Rajaratnam et al., 2008; Khare et al., 2015). In ongoing research, we are also developing methods for model selection that automatically find the graph $G$ using frequentist and Bayesian approaches.

Here, some strategies to define the graph $G$ bringing biological information were proposed; however, these were not implemented in the real data analyses because they imply a formidable challenge at the bioinformatics and biochemical level which is beyond the scope of this study. Therefore, evaluating the performance of our models when $G$ is defined through the use of biological information is another topic for further research. The use of biological information proposed here involves the use of bioinformatics tools to group markers based on features such as their annotated function, i.e., using gene annotation information.

The methods presented in Gianola et al. (2003) assumed a spatial correlation depending on distance between markers, considered only within chromosome correlations and some of them required equally spaced markers. On the other hand, the methods of Yang and Templeman (2012) assumed a non-stationary correlation structure based on the spatial position of the markers (using physical or linkage maps). The methods proposed here are more versatile and flexible because they allow defining the partial correlation structure based on criteria different to physical position. In fact, our methods are a way to incorporate biological information (e.g., knowledge of biochemical pathways.) into statistical models for genome-wide prediction through the use of such information to define the conditional covariance structure of marker allele substitution effects. Moreover, our models can take into account the fact that partial correlation of loci effects in different chromosomes may exist. When using biological information to define $G$, markers are clustered in "functional sets" that were referred to as blocks. This can be done as described in Peñagaricano et al. (2013). Briefly, in a first step they assigned markers to genes and then they used existing databases (e.g., the Gene Ontology database) to define functional classes of genes. Thus, markers in the same functional class conform a block. For further details, see Peñagaricano et al., (2013). One potential limitation of this strategy is that annotation for all markers in a given chip may not be available; consequently, not all markers can be grouped. In this case, an easy solution is to assign these markers without information to a functional group based on physical distance as in Peñagaricano et al. (2013) or some LD metric. Alternatively, the effect of each marker in this set could be considered to be conditionally independent of the effects of the remaining markers. When allowing effects of markers in different blocks to be correlated, conditions for the induced graph to be decomposable are stated in proposition 1. The importance of this proposition is that these conditions can be seen as way to construct a decomposable graph under approach 3.b. This proposition requires the existence of an ordering that coupled with the edges set satisfy conditions 1.1 to 1.4. An ideal scenario would be the one where there exists an ordering of the markers that satisfies these conditions without modifying the edges set. However, if it does not exist, notice that it might be possible to satisfy these conditions by adding and deleting edges. Of course, it is not optimal because it implies removing important edges (i.e., partial covariances) and adding others that may not exist in the true underlying graph. Because it is not desirable to remove edges established on the basis of biological information, it might be better to add edges only. This would increase the number of covariance parameters in the model, but this is the price paid for finding a decomposable graph in order to take advantage of the theoretical and computational advantages of this sort of graphs. Moreover, this can be seen as a compromise between the generality of allowing arbitrary partial correlations between effects of markers in different blocks and the case of conditionally independent blocks (i.e., no conditional covariances among effects of markers in different blocks). Notice that in the independent blocks case, the induced graph is also decomposable, but allowing correlations among marker effects in different blocks while maintaining the advantages of decomposability allows more generality in the modeling process and is more appealing from the biological perspective. Because of these reasons, we believe that this proposition could be useful.

For $m$ variables, there are $m!$ possible orderings which may make the search for an ordering of markers satisfying conditions 1.1 to 1.3 intractable. Therefore, proposition 2 was developed in order to reduce the problem. Using this strategy, a perfect elimination ordering can be found just by manipulating blocks instead of individual markers. Thus, the task consists in finding an ordering of blocks satisfying condition 2.2 (provided the edges set satisfies 2.1 and 1.4). If such an order does not exist, then extra links between blocks could be added to satisfy this condition. However, this strategy is more restrictive in terms of the conditional covariance structure, as imposed by condition 2.1. This is the cost of having an easier way to find a perfect elimination ordering (whose existence implies decomposability). Under corollary to proposition 2, there are scenarios where it is very easy to order markers according to a perfect elimination scheme. For example, when each block is linked at most with one block, or when all blocks are linked, the order of blocks is arbitrary.

The idea of adding edges to a given graph $G$ in order to find a decomposable graph containing it has been used in the realm of graphical models for covariance estimation; this decomposable super-graph is known as the cover of $G$ (Khare and Rajaratnam, 2012). However, the problem of finding this graph is often computationally intractable and there are no general rules on how to proceed. Propositions 1 and 2, and the corollary to proposition 2 provide conditions ensuring decomposability for the particular case described in approach 3.b and also provide rules to build decomposable graphs. In cases where assessing all possible orderings is not possible or there is no ordering that satisfies these conditions, they may guide the process of adding edges, i.e., may indicate what edges have to be added. Moreover, notice that conditions in proposition 2 are easier to assess than those in proposition 1 and that corollary to proposition 2 provides even simpler ways to find a perfect elimination ordering. Once more, notice that the price paid for easing the ordering of markers is imposing a more restrictive covariance structure. However, imposing highly restrictive structures has resulted in some improvements in accuracy. In Yang and Templeman (2012), it was assumed that the effect of each marker was correlated only with the effects of adjacent markers according to the physical position, which is clearly a very restrictive assumption, yet they found some gains in accuracy as discussed before.

## 3.4 Multiallelic loci

Currently, there exists a widespread use of SNP chips to bring genomic information into the statistical models used to carry out prediction of genetic values and phenotypes in animals, humans and plants. These markers are used just as a proxy for the actual genes affecting the trait or set of traits under consideration and they are typically biallelic. However, advances in bioinformatics are allowing the determination of genotypes for actual genes. If all genes of a given species were known, predictions could be based on genotypes of these genes instead of marker genotypes, it has an obvious biological advantage because it is better to work with actual genes in lieu of proxies. Hopefully, such an ideal scenario is going to be attained at some point in the future. In this case, for many loci there would be more than two alleles. Another case in which multiallelic loci have to be considered is the following. Genome-wide prediction models have also considered diplotypes defined from haplotypes that have been built from two or more adjacent markers as explanatory variables (Meuwiseen et al., 2001; Calus et al., 2008); consequently, the number of haplotypes is greater than two. Of course, if the explanatory variables are built based on molecular markers, haplotypes, or actual genes, there are biological implications. Notwithstanding, from a merely statistical point of view, the methods proposed here can be easily extended to the multi-allelic case. If there are $a_h$ alleles at locus $h$, then the corresponding columns of the design matrix are built by defining $a_h - 1$ variables as follows:

$$W^h = \{w_{ij}^h\} = \begin{cases} 1, if\ genotype = A_j A_j \\ 0, if\ genotype = A_j - \ , j = 1, 2, \ldots, a_h - 1, \\ -1, if\ genotype = - - \end{cases}$$

where $w_{ij}^h$ is the "centered" gene content (i.e., gene content-1) of the $i^{th}$ individual for the $j^{th}$ allele and $"-"$ represents an allele different from $A_j$. Across loci, the graph $G$ can be defined using approaches presented in subsection 2.4, but now it has to be considered that there are two or more effects per locus; consequently, the conditional covariance structure among effects of alleles at the same locus has to be defined. Perhaps, a simple approach is to assume that either, all effects at the same locus are partially correlated or that they have a null partial correlation.

## 4. CONCLUSIONS

An important contribution of this paper is the introduction of GCGM in the context of genome-wide prediction to account for partially correlated marker effects, a phenomenon that was known to happen since early stages of this technology and that has not been widely studied. Methods to incorporate this rich and useful theory in the problem of predicting genetic values using genomic data via linear models were developed. Our methods are flexible in the sense that they offer frequentist and Bayesian approaches, allow incorporating biological information in the prediction problem, and permit to model a wide range of conditional covariance structures resulting from different factors (not only markers' position in the genome). One possible way to make our methods more effective is through the use bioinformatics, biochemistry and physiology to build more reliable graphs which may significantly increase the accuracy of genome-wide prediction.

**Author contributions**

C.A. Martínez developed the methods, proved the propositions and the corollary, wrote most of the R scripts, designed and carried out the simulations and wrote the paper. K. Khare advised the development of methods, reviewed and discussed the proofs and statistical aspects of the paper. S. Rahman discussed the methods and wrote some of the R scripts. M.A. Elzo helped in designing the simulation, reviewed and discussed the genetic aspects of the paper.

**Acknowledgements**

C.A. Martínez thanks Fulbright Colombia and "Departamento Administrativo de Ciencia, Tecnología e Innovación" COLCIENCIAS for supporting his PhD and Master programs at the University of Florida through a scholarship, and Bibiana Coy for her love, support and constant encouragement.

## APPENDIX A: BASIC CONCEPTS IN GRAPH THEORY

**Undirected graph.** An undirected graph $G$ is defined as a collection of two objects $G = (V, E)$ where $V$ is the set of vertices (finite) and $E \subseteq V \times V$ is the set of edges satisfying:
$$(u, v) \in E \Leftrightarrow (v, u) \in E.$$
**Neighbor vertices**. Let $G = (V, E)$ be an undirected graph. The vertices $u, v \in V$ are said to be neighbors if $(u, v) \in E$.

**P-path**. A p-path is a collection of p distinct vertices $u_1, u_2, \ldots, u_p$ such that $(u_i, u_{i+1}) \in E, i = 1, 2, \ldots, p-1$, that is, $(u_i, u_{i+1})$ are neighbors for $i = 1, 2, \ldots, p-1$.

**P-cycle**. A p-cycle is a collection of p distinct vertices $u_1, u_2, \ldots, u_p$ such that $(u_i, u_{i+1}) \in E, i = 1, 2, \ldots, p-1$ and $(u_p, u_1) \in E$

**Clique**. A subset $V_0 \subset V$ is a clique if $(u, v) \in E \ \forall \ u, v \in V_0$.

**Maximal clique**. A subset $V_0 \subset V$ is defined to be a maximal clique if $V_0$ is a clique and there does not exist a clique $\bar{V}$ such that $V_0 \subset \bar{V} \subseteq V$.

**Ordered graphs**. Let $G = (V, E)$ and let $\sigma$ be an ordering of $V$, that is, a bijection from $V$ to $\{1, 2, \ldots, |V|\}$. Then, the ordered graph $G_\sigma = (\{1, 2, \ldots, |V|\}, E_\sigma)$ is defined as follows: $(i, j) \in E_\sigma$ iff $(\sigma^{-1}(i), \sigma^{-1}, (j)) \in E$.

**Perfect elimination ordering**. An ordering $\sigma$ of a graph $G = (V, E)$ is defined to be a perfect elimination ordering if a triplet $\{i, j, k\}$ with $i > j > k$ such that $(i, j) \notin E_\sigma$ and $(i, k), (j, k) \in E_\sigma$ does not exist.

**Subgraph.** The graph $G' = (V', E')$ is said to be a subgraph of graph $G = (V, E)$ if $V' \subseteq V$ and $E' \subseteq E$.

**Induced subgraph.** Consider the graph $G = (V, E)$ and a subset $A \subseteq V$. Define $E_A = (A \times A) \cap E$. The subgraph $G_A = (A, E_A)$ is defined to be a subgraph of $G$ induced by $A$.

**Decomposable graph**. This is an important concept in graphical models. A graph $G = (V, E)$ is a decomposable graph if it does not contain a cycle of length greater than or equal to four as an induced subgraph. It turns out that decomposable graphs are characterized by the existence of a perfect elimination ordering of their vertices; therefore, a graph $G = (V, E)$ is decomposable iff its vertices admit a perfect elimination ordering.

## APPENDIX B: PROOF OF PROPOSITION 1

Approach 1 induces either a banded or a differentially banded matrix and these structures are known to correspond to decomposable graphs. Similarly, 3.a induces block diagonal matrices which also correspond to decomposable graphs. The only non-trivial part is to prove the statements made about conditions for 3.b to induce a decomposable graph. To prove it, we show that it is possible to find a perfect elimination ordering, and we use the fact that the existence of a perfect elimination ordering characterizes decomposable graphs.

**Proof of sufficiency**

Hereinafter, consider arbitrary indices $\{i, j, k\}$ such that according to $\sigma'$ $i > j > k$.

**Case 1**) The triplet $\{i, j, k\}$ pertains to the same block. Because each block corresponds to a complete graph, it follows that in this case it cannot be that

$$i > j > k, (i,j) \notin E_\sigma, (i,k), (j,k) \in E_\sigma \qquad (A.1)$$

**Case 2**) $i \in l, \{j, k\} \in l', (l, l') \in \mathcal{L}$. If $(i, j) \notin E_\sigma$, by condition 1.4 it means that one of the following mutually exclusive events must occur:

- $i \notin C_{l_{l'}}$ and $j \in C_{l'_l}$
- $i \in C_{l_{l'}}$ and $j \notin C_{l'_l}$
- $i \notin C_{l_{l'}}$ and $j \notin C_{l'_l}$.

It follows that:

$$i \notin C_{l_{l'}} \Rightarrow (i, k) \notin E_\sigma$$

this suffices to show that condition $A.1$ cannot be attained under the first and third scenarios. On the other hand:

$$i \in C_{l_{l'}} \text{ and } j \notin C_{l'_l} \Rightarrow k \notin C_{l'_l} \text{ (}\because \text{ condition 1.3)}$$
$$\Rightarrow (i, k) \notin E_\sigma,$$

which is enough to show that condition $A.1$ does not hold under the second scenario.

**Case 3)** $(i,j) \in l, k \in l'$. In this case $(i,j) \in E_\sigma$ because all blocks are complete; therefore, condition $A.1$ cannot be met.

**Case 4)** $i \in l, j \in l', k \in l''$. If $k \notin C_{l''_l} \cup C_{l''_{l'}}$, it is clear that $(j,k), (i,k) \notin E_\sigma \; \forall \, i \in B_l, \forall \, j \in B_{l'}$. If each one of the three blocks is linked with the other two, that is, $(l,l'), (l',l''), (l,l'') \in \mathcal{L}$, condition 1.1 implies that if all three intersection sets are non-empty and $(i,j) \notin E_\sigma$ then either $(i,k) \notin E_\sigma$ or $(j,k) \notin E_\sigma$ which prevents the occurrence of $A.1$. If one or two of the three intersection subsets is empty, by condition 1.1 it follows that $I_{l''} = \emptyset$ then:

$$k \in C_{l''_l} \Rightarrow (j,k) \notin E_\sigma$$
$$k \in C_{l''_{l'}} \Rightarrow (i,k) \notin E_\sigma,$$

Notice that this holds disregarding the location of indices $j$ and $i$ within their corresponding blocks; therefore, condition $A.1$ cannot be attained.

On the other hand, if one block is linked to the other two, and these are the only existent links, condition 1.2 guarantees that $A.1$ does not hold.

Until here, only linked blocks were considered. Notice that if at least one index of the triplet $\{i,j,k\}$ corresponds to a marker in an isolated (i.e. not linked) block, then $A.1$ does not happen.

In conclusion, under the four conditions of result 1 it is possible to find a perfect elimination ordering and therefore the graph induced under approach 3.b is decomposable.

Now we proceed to prove necessity of conditions 1, 2 and 3. To this end, it is proven that under 3.b, if these conditions do not hold, then the induced graph is not decomposable.

**Proof of necessity of conditions 1.1, 1.2 and 1.3**

Hereinafter when the words "there is always" precede a condition, it is meant that under any ordering of markers, the graph always satisfies the condition.

Suppose that 1.1 does not hold. It means that:

- There is always at least one triplet of linked blocks $\{l, l', l''\}$, $1 \leq l, l', l'' \leq |\Psi|$, linked in the following way: $C_{l_{l'}} \neq C_{l_{l''}}$, $C_{l'_l} \neq C_{l'_{l''}}$, $C_{l''_l} \neq C_{l''_{l'}}$, the three pairwise intersections $I_l$, $I_{l'}$ and $I_{l''}$ are non-empty, and there exist at least one triplet $\{i,j,k\}$ such that $i > j > k$ and $i \in C_{l_{l''}} \cap I_l^c, j \in C_{l'_l}$ or $i \in C_{l_{l''}}, j \in C_{l'_l} \cap I_{l'}^c$, and $k \in I_{l''}, i > j > k$,

or

- There is always at least one triplet of linked blocks $\{l, l', l''\}$ and a triplet of indices $\{i,j,k\}$ such that $i > j > k$, exactly one or two of the three sets $\{I_l, I_{l'}, I_{l''}\}$ are empty and $k$ pertains to a non-empty intersection set.

It is easy to notice that in any of the two events described in the first case, condition A.1 is satisfied. On the other hand, in the second case, if $k$ pertains to the (non-empty) intersection set of a block, say $I_l$, then we can always pick $i \in C_{l''_l}$ and $j \in C_{l'_l}$ such that $(i,j) \notin E_\sigma$, then, because $k \in I_l$ it follows that $(i,k), (j,k) \in E_\sigma$, that is, $A.1$ holds.

Therefore, if 1.1 does not hold, a perfect elimination ordering does not exist and the graph is not decomposable. Similarly, if there is always at least one triplet of blocks $\{l, l', l''\}$ satisfying the converse of condition 1.2 then $A.1$ always holds, that is, a perfect elimination ordering does not exist and consequently the graph is not decomposable. Finally, if 1.3 does not hold then always exists at least a linked pair $\{l, l'\}$ such that $\exists\, i \in C_{l_{l'}}, j \in C^C_{l'_l} \cap l, k \in C_{l'_l}, i > j > k$ which immediately implies $A.1$.

**Proof of non-necessity of condition 1.4**

We refute the statement that condition 1.4 is necessary by constructing a counter-example. Suppose that a graph is comprised of two blocks. Let $L_\sigma$ be the subset of $E_\sigma$ containing edges comprised of one node pertaining to block 1 and one node from block 2. Also assume that block 1 contains nodes $\{a, b, c, d\}$, $C_{1_2} \coloneqq \{b, c, d\}, C_{2_1} \coloneqq \{w, z\}$ and $\mathcal{L} \coloneqq \{(d, w), (c, w), (b, z), (c, z), (d, z)\}$. Consider any ordering satisfying the following: $\sigma(a) = 1, \sigma(b) = 2$, $\sigma(c) = 3, \sigma(d) = 4$. Consider $i > j > k, i \in C_{2_1}$, if $i = \sigma(w)$ then $(i, j) \notin E_\sigma \Rightarrow (i, k) \notin E_\sigma$, if $i = \sigma(z)$ then $(i, j) \notin E_\sigma \Rightarrow j = 1 = \sigma(a)$ $\therefore \nexists\, k < j$ and consequently $A.1$ does not hold. We have found a family of perfect elimination orderings and its existence implies that the graph is decomposable even though condition 4 does not hold. It proves that condition 3 is not a necessary condition.

∎

## Appendix C: Proof of proposition 2

**Proof of sufficiency** The general approach is very similar to that used to prove sufficiency of conditions stated in result 1. Let $\Lambda$ be the set of blocks linked to at least one block. Assume that markers have been ordered as stated in result 2, this ordering is denoted by $\sigma$. Let $B_l^\sigma$ be the set of indices corresponding to markers in block $l$ under $\sigma$. Let $\{i, j, k\}$ be three arbitrary indices such that according to $\sigma$, $i > j > k$, and $C_l \in B_l^\sigma$, $l \in \Lambda$ be the subset of indices corresponding to markers in block $l$ whose effects are partially correlated with the effects of a subset of markers in another block $1 \leq l \leq |\Lambda|$. In the following, it is proven that under conditions 1.4, 2.1 and 2.2 this within-block ordering strategy yields a perfect elimination ordering. Four cases depending on the position of indices $i, j, k$ are considered.

**Case 1)** The triplet $\{i, j, k\}$ pertains to the same block. Because each block corresponds to a complete graph, it follows that in this case condition $A.1$ is not satisfied.

**Case 2)** $i \in l, \{j, k\} \in l', \{l, l'\} \in \Lambda, \rho(l) > \rho(l')$. If $(i, j) \notin E_\sigma$, by condition 1.4, it means that one of the following mutually exclusive events must occur:

- $i \notin C_l$ and $j \in C_{l'}$
- $i \notin C_l$ and $j \notin C_{l'}$
- $i \in C_l$ and $j \notin C_{l'}$

It follows that:
$$i \notin C_l \Rightarrow (i, k) \notin E_\sigma,$$
this suffices to show that condition $A.1$ cannot be attained under the first and second events. On the other hand:
$$i \in C_l \text{ and } j \notin C_{l'} \Rightarrow k \notin C_{l'} \ (\because k < j \text{ under } \sigma)$$
$$\Rightarrow (i, k) \notin E_\sigma,$$
which is enough to show that condition $A.1$ does not hold under the second scenario.

**Case 3)** $\{i,j\} \in l, k \in l', l > l'$. In this case $(i,j) \in E_\sigma$ because all blocks are complete; therefore, condition $A.1$ cannot be met.

**Case 4)** $i \in l, j \in l', k \in l'', \rho(l) > \rho(l') > \rho(l'')$. If $k \notin C_{l''}$ it is clear that $(j,k), (i,k) \notin E_\sigma \; \forall \, i \in B_l \; \forall \, j \in B_{l'}$ disregarding the position of indices $i$ and $j$ within their corresponding blocks and the linkage relationship among the three blocks, thus $A.1$ does not hold and, therefore; in the following, only the case $k \in C_{l''}$ is considered. If each one of the three blocks is linked with the other two, then:

$$(i,j) \notin E_\sigma \Rightarrow i \notin C_l \text{ or } j \notin C_{l'}$$
$$\Rightarrow (i,k) \notin E_\sigma \text{ or } (j,k) \notin E_\sigma \; (\because \text{condition 2.1}).$$

Similarly, if one block is linked to the other two, and these are the only existent links, condition 2.2 guarantees that $A.1$ does not hold.

Consequently, under the two conditions of result 2 and condition 1.4, ordering $\sigma$ is a perfect elimination ordering. Thus, the graph induced under approach 3.b with and edges set as defined in result 2 is decomposable.

**Proof of necessity of condition 2.2**

To prove necessity of condition 2.2, we show that if it is not satisfied, then $\sigma$ is not a perfect elimination ordering. If there is at least one triplet of blocks $\{l, l', l''\}$ such that $(l, l'), (l, l'') \in \mathcal{L}, (l', l'') \notin \mathcal{L}, \rho(l) < \rho(l') < \rho(l'')$ then by picking $i \in C_{l''}^l, j \in C_{l'}^l$ and $k \in C_{l'}^l$, it follows that $i > j > k$, and $(i,j) \notin E_\sigma, (i,k), (j,k) \in E_\sigma$, that is, condition A.1 holds and as a consequence, $\sigma$ is not a perfect elimination ordering, which implies necessity of condition 2.2.

**Proof of non-necessity of condition 2.1**

By constructing a counter example, we refute the necessity of condition 2.1. Suppose that $G$ contains three blocks such that $\rho(B_1) = 1, \rho(B_2) = 2, \rho(B_3) = 3$, $B_1$ is only linked to $B_2$ and $B_2$ is only linked to $B_3$, $C_{2_1} \neq C_{2_3}$ and markers have been ordered according to $\sigma$. It is easy to see that in this graph, for any triplet $\{i,j,k\}$ such that according to $\sigma, i > j > k$ if $(i,j) \notin E_\sigma$ then $(i,k) \notin E_\sigma$ which implies that condition A.1 is never reached and consequently, $\sigma$ is a perfect elimination ordering despite of the fact that condition 2.1 does not hold. Consequently, 2.1 is not a necessary condition.

∎

## Appendix D: Proof of corollary to proposition 2

Let $G^s$ be the "super graph" defined in the corollary. It is assumed that conditions 2.1 and 1.4 hold. Consequently, we only need to check that condition 2.2 is satisfied. It follows immediately by noticing that condition 2.2 is nothing but the definition of a perfect elimination ordering of the blocks, i.e., a perfect elimination ordering of the "super vertices" of $G^s$.

∎